\documentclass{moriond}

\def\be{\begin{equation}}
\def\ee{\end{equation}}
\def\bea{\begin{eqnarray}}
\def\eea{\end{eqnarray}}

\newcommand{\Photo}{}

\newcommand{\ctsper}      {cts/(keV$\cdot$kg$\cdot$yr)}

\newcommand{\kgy}         {{kg$\cdot$yr}}

\newcommand{\qbb}         {{$Q_{\beta\beta}$}}

\newcommand{\bb}        {{$\beta\beta$}}

\newcommand{\onbb}        {{$0\nu\beta\beta$}}

\newcommand{\gerda}       {\textsc{Gerda}}
\newcommand{\majorana}       {\textsc{Majorana}}
\newcommand{\legend}       {LEGEND}

\newcommand{\gess}        {$^{76}$Ge}
\newcommand{\tony}        {{ton$\cdot$yr}}
\newcommand{\ctsfwhm}      {cts/(FWHM$\cdot$ton$\cdot$yr)}

\graphicspath{{pics/}}
\usepackage[pagewise]{lineno}

\begin{document}
\vspace*{4cm}
\title{NEUTRINOLESS DOUBLE BETA DECAY SEARCH WITH $^{76}$Ge:\\STATUS AND PROSPECT WITH LEGEND}

\author{V. D'Andrea for the \legend~Collaboration}

\address{Dipartimento di Scienze Fisiche e Chimiche, Universit\`a dell'Aquila, Italy \footnote{also at INFN Laboratori Nazionali del Gran Sasso}}

\maketitle\abstracts{The search for neutrinoless double beta ($0\nu\beta\beta$) decay is the best way to test lepton number violation and Majorana nature of neutrinos. One of the most promising techniques to discover $0\nu\beta\beta$ decay is by operating High-Purity Ge detectors enriched in $^{76}$Ge. The current generation of $^{76}$Ge experiments, \textsc{Gerda} and \textsc{Majorana}, lead the field in the achieved energy resolution and ultra-low background. These are two of the most important characteristics for sensitive searches of this undiscovered decay. The next generation of $0\nu\beta\beta$ decay experiments requires more mass and further reduction of backgrounds to maximize the discovery potential. Building on the successes of \textsc{Gerda} and \textsc{Majorana}, the LEGEND collaboration has been formed to pursue a tonne-scale $^{76}$Ge experiment, with a discovery potential projected to be a half-life beyond $10^{28}$~years. The collaboration aims to develop a phased experimental program, starting with a 200~kg measurement by repurposing the existing \textsc{Gerda} infrastructure.}

\section{Introduction}
The dominance of the matter over the antimatter in our universe is one of the most interesting aspects of cosmology. One of the favored models to explain this dominance is the leptogenesis~\cite{leptogenesi}, that is based on the violation of the lepton number.
In many extensions of the Standard Model~\cite{numass}, neutrinos are assumed to be their own antiparticles (Majorana particles), explaining the origin of the low neutrino mass and leading to lepton number violating processes.
At present, the only feasible experiments having the potential of establishing that the massive neutrinos are Majorana particles are the ones searching for the neutrinoless double beta ($0\nu\beta\beta$) decay. 

\section{Search for neutrinoless double beta decay}
The double beta ($\beta\beta$) decay is a second order weak nuclear decay process with extremely long half-life, consisting of the transformation of a pair of neutrons into two protons as a single process with the emission of two electrons. The standard model predicts the $\beta\beta$ decay with two neutrinos ($2\nu\beta\beta$):
$(Z, A) \rightarrow (Z + 2, A) + 2e + 2\bar{\nu}_e$, this decay has been observed in a few isotopes.

The neutrinoless mode of this decay is not predicted by the Standard Model and consists of the emission of only two electrons: $(Z, A) \rightarrow (Z + 2, A) + 2e$. This decays violates the lepton number conservation by two units and has never been observed up to now.

The search for a $0\nu\beta\beta$ decay signal consists of the detection of the two emitted electrons, with total energy corresponding to the mass difference \qbb~of the two nuclei. 
The rate of the $0\nu\beta\beta$ decay is usually factorized into three terms~\cite{doi}:
\begin{equation}
 \left(T_{1/2}^{0\nu}\right)^{-1}=G_{0\nu}|M_{0\nu}|^2 \left(\frac{m_{\beta\beta}}{m_e}\right)^2
 \label{Eq:rate_0nu}
\end{equation}
where $T_{1/2}^{0\nu}$ is the half-life of the $0\nu\beta\beta$ process, $G_{0\nu}$ is the phase space factor (PSF) and $M_{0\nu}$ is the nuclear matrix element (NME) \cite{iachello}.
In the expression of Eq.~(\ref{Eq:rate_0nu}) a fundamental quantity appears, the effective Majorana mass $m_{\beta\beta}=|\sum_{i=1}^3 U_{ei}^2 m_i|$
(where $U$ is the PMNS mixing matrix and $m_i$ are the neutrino mass eigenvalues). 
The key idea of the experiments is that, by studying the $0\nu\beta\beta$ decay, it is possible to measure its half-life and then estimate $m_{\beta\beta}$.

The sensitivity of a given  experiment is expressed by~\cite{delloro}:
\begin{equation}\label{Eq:sensNu}
 S^{0\nu} = \frac{\ln 2 \cdot N_A \cdot \epsilon \cdot f_{ab}}{m_A}\cdot \frac{1}{n_\sigma} \cdot \sqrt{\frac{M \cdot T}{BI \cdot \Delta E}}~.
\end{equation}
This formula emphasizes the role of the experimental parameters needed in the search of the  decay: the detection efficiency $\epsilon$, the isotopic abundance $f_{ab}$ of the $\beta\beta$ emitter, the target mass $M$, the experimental live-time $T$, the background index $BI$ and the energy resolution $\Delta E$.

Of particular interest is the case in which $BI$ is so low that the expected number of background events is less than one count within the energy region of interest ($Q_{\beta\beta} \pm 0.5$~full-width at half-maximum, FWHM) and a given exposure: this is called ``background-free'' condition. Next generation experiments aim for having this condition. The first data release after the upgrade~\cite{nature} showed that \textsc{Gerda} is the first background-free experiment in the  field, since it will remain in this condition up to its design exposure. The advantage of this condition is that the sensitivity $S^{0\nu}$ grows linearly with the experimental mass and time, instead of by square root like in Eq.~(\ref{Eq:sensNu}).

The most recent results on $0\nu\beta\beta$ decay, including half-life lower limits and sensitivities and corresponding sensitivity ranges on the effective Majorana mass $m_{\beta\beta}$ are listed in Tab.~\ref{Tab:limit}.


\begin{table}[h]
\caption{\label{Tab:limit} Results from different \onbb~decay experiments: lower half-life limits $T^{0\nu}_{1/2}$ and sensitivities $S^{0\nu}$ (both at 90\% C.L.). The sensitivities $S^{0\nu}$ have been converted into upper limits of effective Majorana masses $m_{\beta\beta}$ with the corresponding NME~\protect\cite{engel},
the ranges are reported in the table.}
\vspace{0.4cm}
\begin{center}
\begin{tabular}{|c|c|c|c|c|}
\hline
isotope  & $T^{0\nu}_{1/2}$ [10$^{25}$~yr] & $S^{0\nu}$ [10$^{25}$~yr] & $m_{\beta\beta}$ [eV] & experiment \\
  \hline
   ${}^{76}$Ge & 9    & 11  & $104$\textendash 228 & \textsc{Gerda} \cite{neutrino2018}\\
   ${}^{76}$Ge & 2.7  & 4.8 & $157$\textendash 346 & \textsc{Majorana} \cite{majorana} \\
  ${}^{130}$Te & 1.5  & 0.7 & $162$\textendash 757 & CUORE \cite{cuore}\\
  ${}^{136}$Xe & 1.8  & 3.7 & $93$\textendash 287 & EXO-200 \cite{exo} \\
  ${}^{136}$Xe & 10.7 & 5.6 & $76$\textendash 234 & KamLAND-Zen \cite{kamland}\\
 \hline
\end{tabular}
\end{center}
\end{table}

\section{Neutrinoless double beta decay search with the $^{76}$Ge isotope}
Several isotopes with different techniques are used to search for \onbb~decay. One of the most promising is the \gess: High-Purity Ge (HPGe) detectors acting both as source and detector are used for this purpose. Experiments using this method were first developed in the 1980s~\cite{Bellotti1986}, obtaining limits on the \onbb~decay half-life of $\sim 10^{23}$~yr. In the 1990s the experiments \textsc{HdM}~\cite{hdm} and \textsc{Igex}~\cite{igex} produced for the first time Ge detectors enriched in \gess: the limit was increased to $\sim 10^{25}$~yr.
Currently two experiments are continuing the search for the \gess~\onbb~decay with different techniques, in order to reach a sensitivity larger than $10^{26}~$yr: the \gerda~experiment and the \textsc{Majorana Demonstrator}, details on these experiments are presented in Secs.~\ref{Sec:gerda} and \ref{Sec:majorana}.
The search for \onbb~decay in \gess~will be continued in the following years by the \legend~experiment: with a staged approach it aims to reach a sensitivity on the \onbb~decay  half-life up to $10^{28}$~yr. The experimental program of \legend~is presented in Sec.~\ref{Sec:legend}.

\section{The GERDA experiment}\label{Sec:gerda}
The \textsc{Gerda} experiment~\cite{gerda2013} is located at the underground Laboratori Nazionali del Gran Sasso (LNGS) of INFN in Italy. A rock overburden of about 3500~m water equivalent removes the hadronic components of cosmic ray showers and reduces the muon flux at the experiment.

\begin{figure}
\begin{minipage}{0.5\linewidth}
 \centerline{\includegraphics[width=\linewidth]{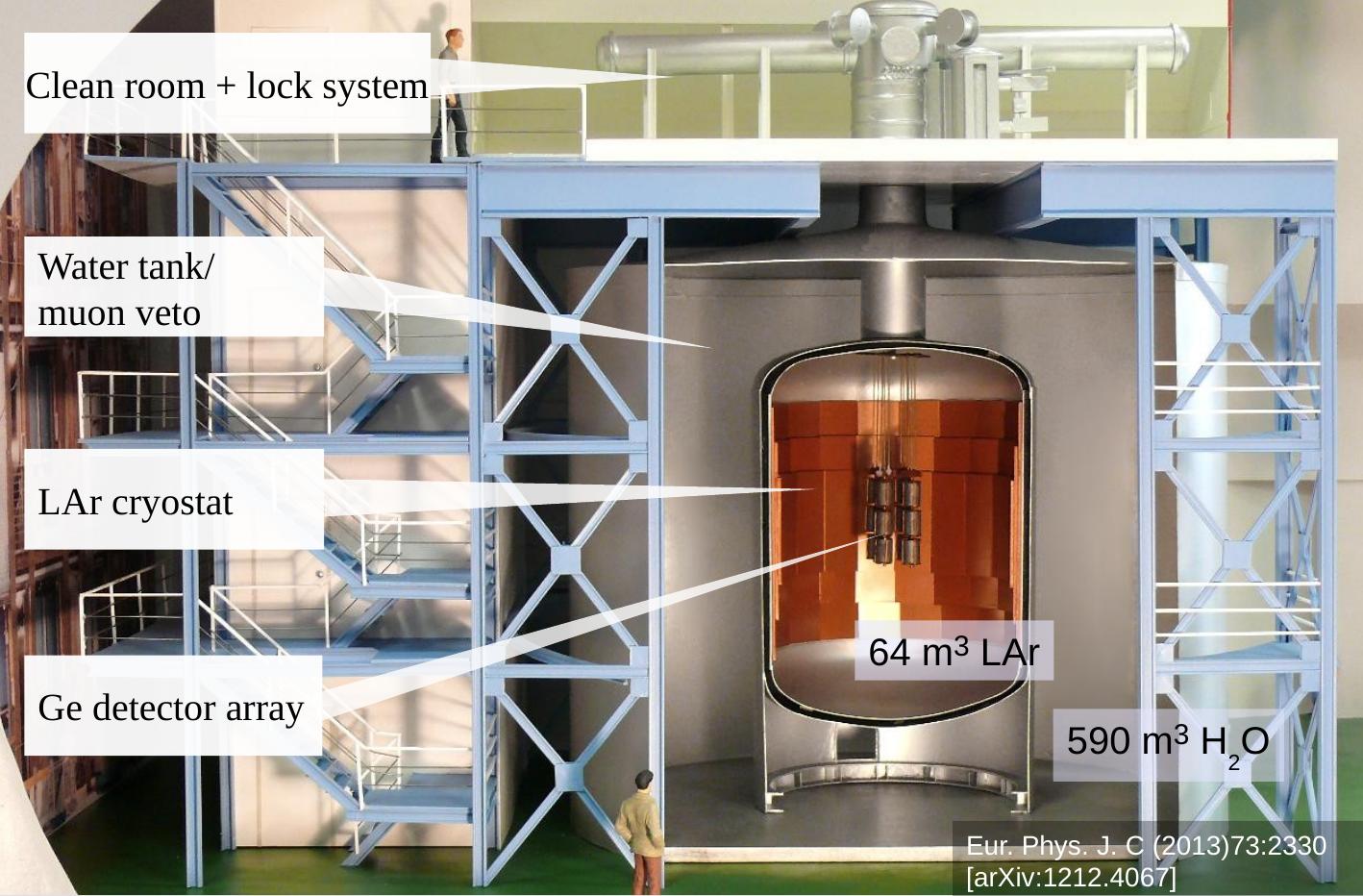}}
 \end{minipage}
 \hfill
\begin{minipage}{0.5\linewidth}
 \centerline{\includegraphics[width=0.6\linewidth]{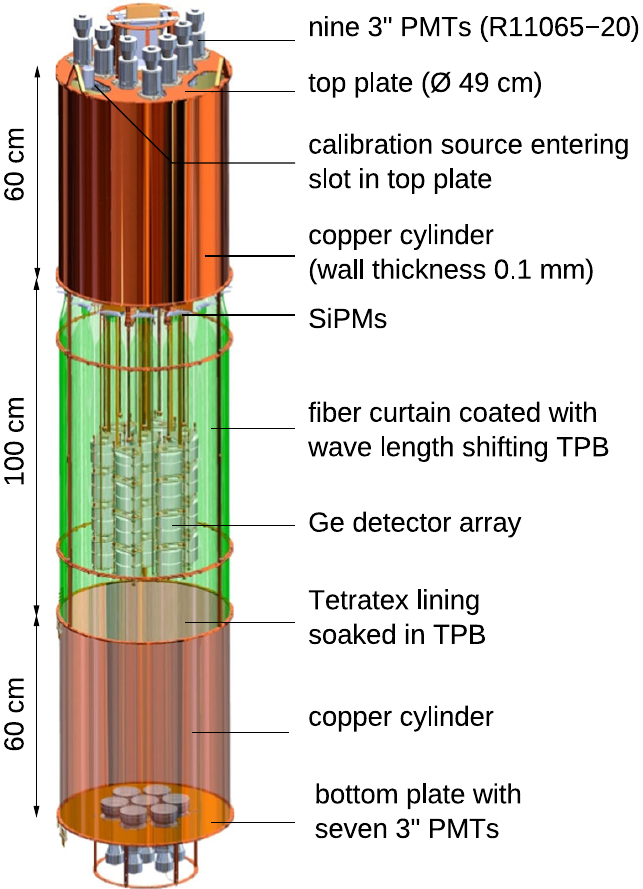}}
 \end{minipage}
 \caption{Left: setup of the \gerda~experiment~\protect\cite{gerda2013}. Right: assembly of detector array and LAr veto system~\protect\cite{upgrade}.}
\label{Fig:view} 
\end{figure}

\paragraph{Design}
The \textsc{Gerda} setup, illustrated in Fig.~\ref{Fig:view} (left), has been designed following a multi-layer approach. HPGe detectors enriched to about 87\% in \gess~are operated bare in liquid argon (LAr). 
The LAr cryostat is complemented by a water tank with 10~m diameter which further shields from neutron and $\gamma$ backgrounds and also works as muon veto.

After a first physics data taking campaign~\cite{prl111} carried out from 2011 to 2013, the \gerda~setup has been upgraded to perform the next step~\cite{upgrade}. The major upgrade is the introduction of 30 new BEGe detectors from Canberra~\cite{bege} with an optimal energy resolution, due to the low input capacitance ($\sim$~pF), and a powerful pulse shape discrimination (PSD), thanks to the configuration of the p$^+$ and n$^+$ contacts that produce a highly non-uniform electrical field.
In addition, an active suppression of the background by detecting the LAr scintillation light, using PMTs and wavelength shifting fibers coupled to SiPMs, has been introduced. 
The core of the \gerda~setup is shown in Fig.~\ref{Fig:view} (right): the Ge detector array (30 BEGe, 7 enriched coaxial and 3 natural coaxial detectors) is at the center of the instrumented LAr volume.

\begin{figure}
\begin{minipage}{0.53\linewidth}
 \centerline{\includegraphics[width=\linewidth]{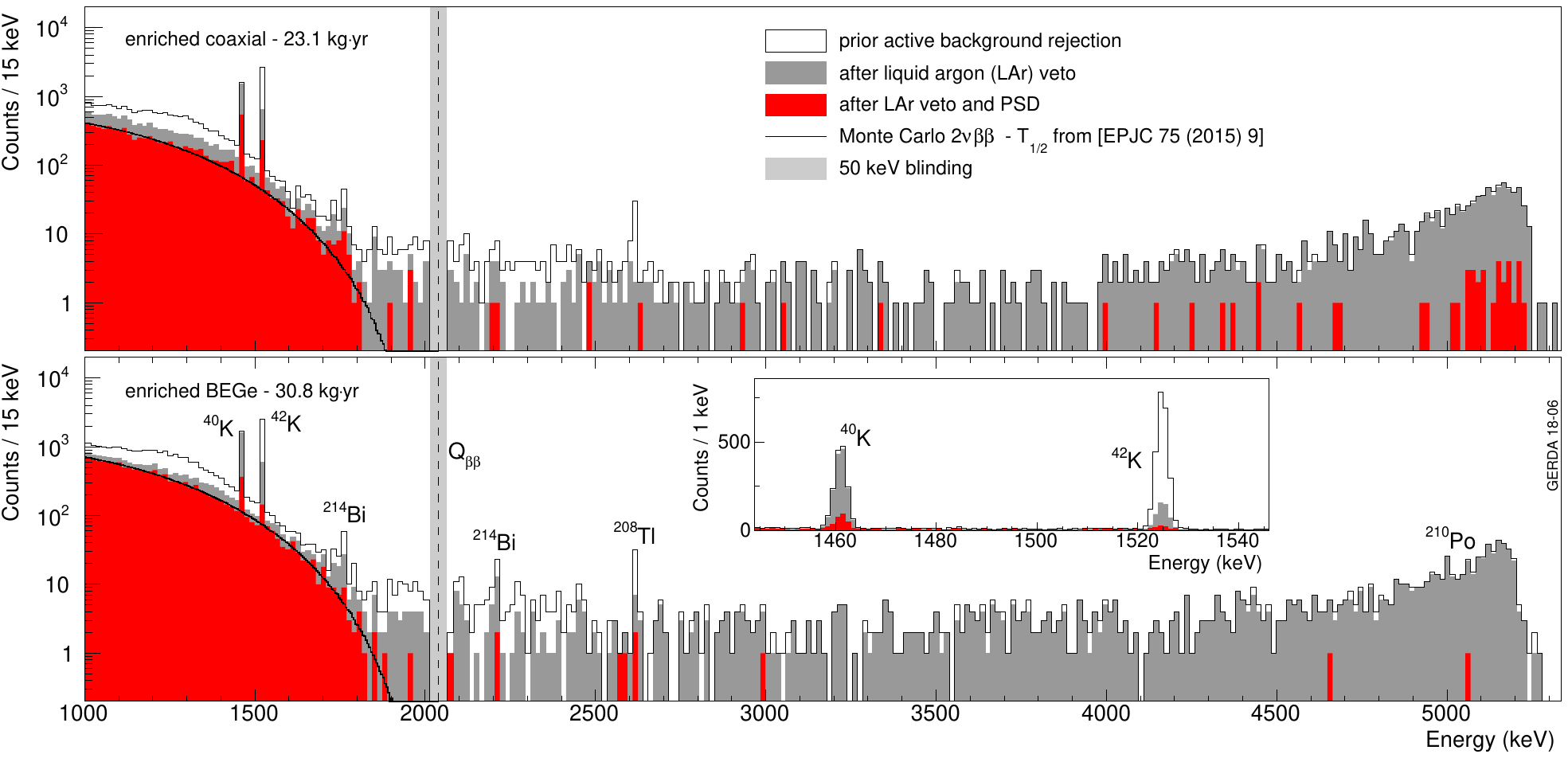}}
 \end{minipage}
 \hfill
\begin{minipage}{0.47\linewidth}
 \centerline{\includegraphics[width=0.9\linewidth]{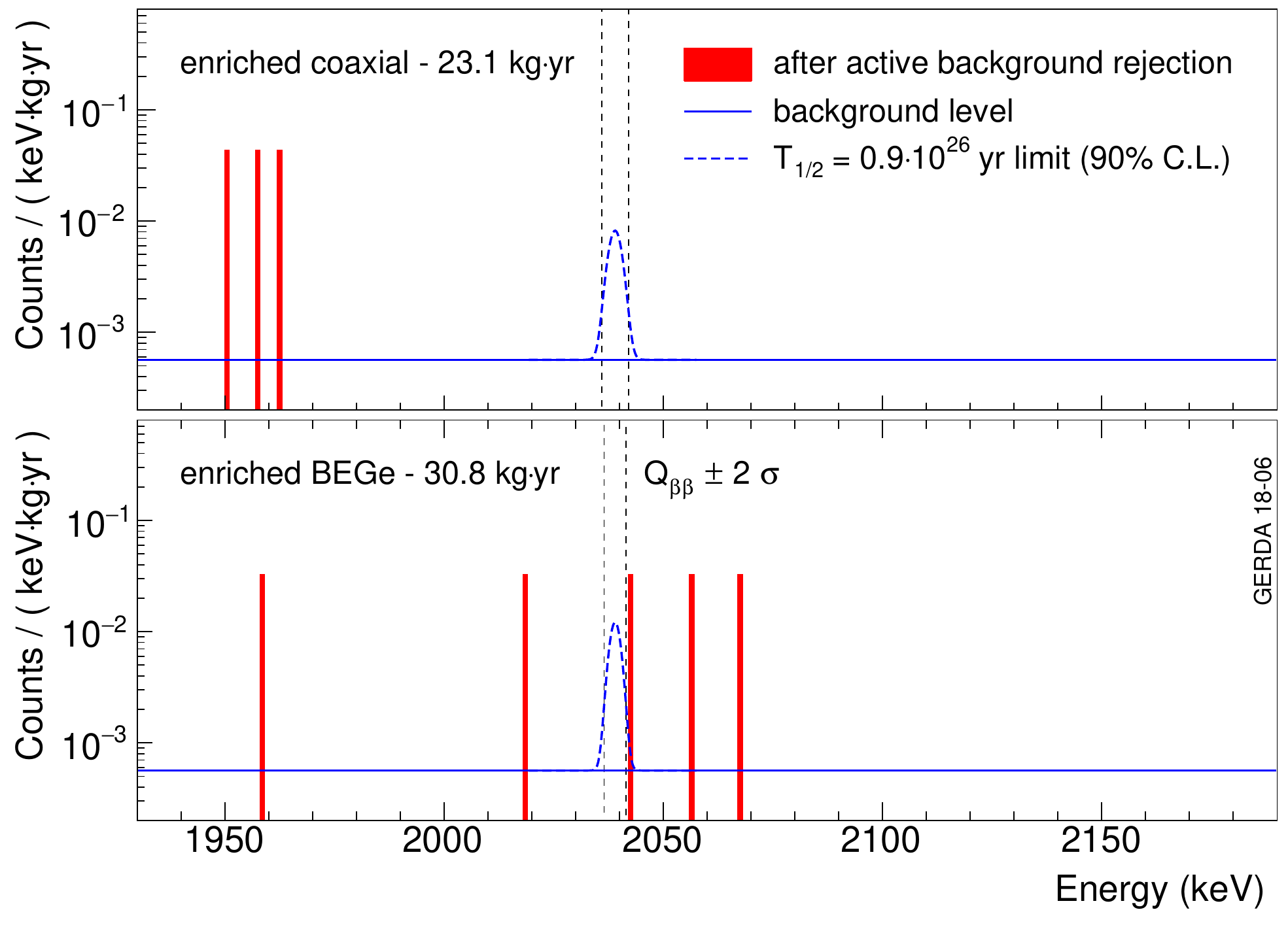}}
 \end{minipage}
 \caption{Left: \textsc{Gerda} energy spectra for enriched coaxial (top panel) and BEGe (bottom panel) detectors before and after the LAr veto and PSD cuts. Right: events observed in the analysis window for coaxial (top) and BEGe (bottom) detectors. The blue lines show fitted background level and the 90\% C.L. limit on \onbb~decay~\protect\cite{neutrino2018}.}
\label{Fig:spectrum} 
\end{figure}

The \textsc{Gerda} background is further reduced by applying PSD cuts~\cite{psdI}. 
For BEGe detectors the PSD is based on the ratio between the peak amplitude of the current signal $A$ and the total energy $E$ ($A/E$): low values are typical for multi-site events ($\gamma$-rays and $\beta$ decays on n$^+$ contacts), high $A/E$ values are from surface events due to $\alpha$ decays on p$^+$ contacts. After the $A/E$ cut, the average survival probability of a $0\nu\beta\beta$ decay event is ($87.6 \pm 2.5$)\%~\cite{neutrino2018}. 
For coaxial detectors the PSD between single-site and multi-site events is based on an artificial neural network (ANN)~\cite{psdI}. 
Additionally, a cut on the risetime of the pulses is applied to reject fast signals from surface events due $\alpha$ decays near the p$^+$ electrode and in the groove. 
The combined PSD efficiency for coaxial detectors is ($71.2\pm4.3$)\%~\cite{neutrino2018}. In Fig.~\ref{Fig:spectrum} (left) the \textsc{Gerda} energy spectra are shown for enriched coaxial (top panel) and BEGe (bottom panel) detectors with the application of LAr veto (in grey) and PSD cuts (in red).

\paragraph{Results}

The total available enriched Ge exposure in \gerda~after the last data release is 82.4~kg$\cdot$yr~\cite{neutrino2018}.
The final spectra in the analysis region are shown in Fig.~\ref{Fig:spectrum} (right): for the coaxial detectors only three events survived, corresponding to a background of $5.7_{-2.6}^{+4.1}\cdot10^{-4}~$cts/(keV$\cdot$kg$\cdot$yr), for BEGe detectors five events remain obtaining a background of $5.6_{-2.4}^{+3.4}\cdot 10^{-4}~$cts/(keV$\cdot$kg$\cdot$yr). With this result \textsc{Gerda} reaches the lowest background ever achieved in the field, taking into account the energy resolution, and will remain in the background-free condition.

The $0\nu\beta\beta$ decay analysis yielded no signal, setting a new limit on the $^{76}$Ge $0\nu\beta\beta$ decay half-life of $T_{1/2}^{0\nu} > 0.9\cdot10^{26}~$yr (90\% C.L.) with a median sensitivity of $1.1\cdot10^{26}~$yr (90\% C.L.), thus making \textsc{Gerda} the first experiment to surpass $10^{26}$~yr sensitivity (as reported in Tab.~\ref{Tab:limit}). 
The fact that the actual $T_{1/2}^{0\nu}$ limit is weaker than the median sensitivity is due to the presence of an event close to $Q_{\beta\beta}$ with energy of 2042.1~keV (2.4~$\sigma$ away from the $Q_{\beta\beta}$).

\section{The MAJORANA DEMONSTRATOR}\label{Sec:majorana}

The \textsc{Majorana Demonstrator}~\cite{majorana2014} is operating an array of HPGe detectors at the Sanford Underground Research Facility (SURF) in Lead, South Dakota with the goal of demonstrating backgrounds low enough to justify construction of a tonne scale Ge based experiment. The array consists of 58 HPGe detectors, with a total Ge mass of 44.8 kg: 14.4 kg of natural Ge detectors and 29.7 kg of detectors enriched to $88.1\pm 0.7\%$ in \gess.

\begin{figure}
 \begin{minipage}{0.5\linewidth}
 \centerline{\includegraphics[width=\linewidth]{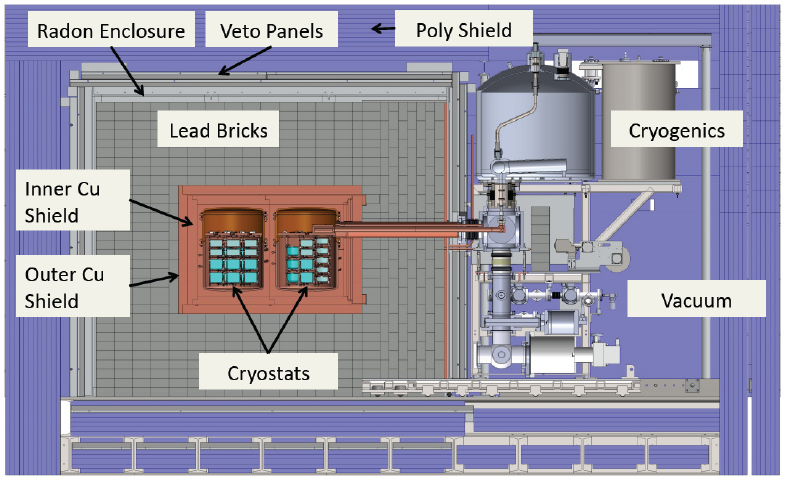}}
 \end{minipage}
 \hfill
 \begin{minipage}{0.5\linewidth}
 \centerline{\includegraphics[width=\linewidth]{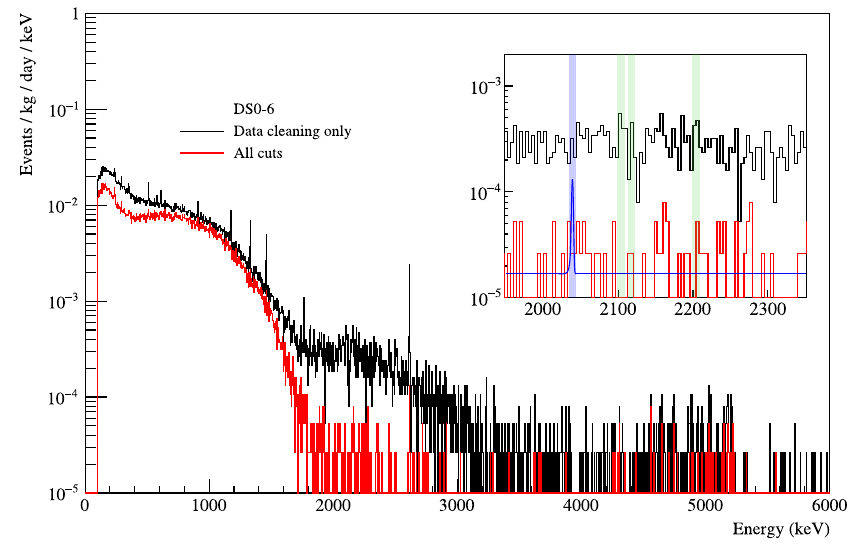}}
 \end{minipage}
\caption[]{Left: design of the \textsc{Majorana Demonstrator}~\protect\cite{majorana2014}. Right: energy spectrum for the full exposure of 26~\kgy~(see the text for more details), the range from 1950-2350~keV is shown in the inset~\protect\cite{majorana}.}
 \label{Fig:majorana}
\end{figure}

\paragraph{Design} The enriched detectors are p-type point contact (PPC) detectors with low capacitance and sub-keV energy thresholds, permitting low-energy physics studies. These detectors have achieved an energy resolution of $2.53 \pm 0.08~$keV (FWHM at $Q_{\beta\beta}=2039$~keV), the best value in the \onbb~decay field.
The experiment utilizes a number of ultra-low activity materials and methods to reduce environmental backgrounds. 
The detectors are split between two modules contained in a low-background copper shield. The copper shielding is contained within 45~cm of high-purity lead shielding, separating the low-background environment from the 
laboratory environment. The lead shield is enclosed within a radon exclusion volume. 
An active muon veto surrounds the radon exclusion volume. 
The design of the \majorana~is shown in Fig.~\ref{Fig:majorana} (left).

To further reduce the background two powerful PSD cuts are applied to the data. PPC detectors have a weighting potential that is relatively low in the bulk of the crystal and peaked in the vicinity of the point contact. This permits to discriminate the $\gamma$-ray background (multi-site events in Ge) from a \bb~decay signal (single-site event) with an efficiency of 90\%~\cite{majorana}. A second PSD is able to discriminate $\alpha$ contaminations. Due to the lithium dead layers on PPC detectors, $\alpha$ particles cannot penetrate in the active region but, impinging on the passivated surface, can deposit energy and can be a potential background near \qbb. However, the slow collection of the holes can be used to discriminate such events from interactions in the crystal bulk. A high efficiency (99.9\%) cut based on the slope of the waveform is implemented~\cite{majorana}.


\paragraph{Results}
The results from the total enriched Ge exposure of 26.0~\kgy~collected in \majorana~until May 2018~\cite{majorana} are presented in the following.
Fig.~\ref{Fig:majorana} (right) shows the measured energy spectra above 100~keV: in black the spectrum with only data cleaning cuts, in red the coincidence between detectors and PSD cuts are also applied.
The inset of Fig.~\ref{Fig:majorana} (right) shows the background spectrum in the energy range from $1950-2350~$keV. After applying all cuts, the background from the resulting 360~keV window is $(6.1 \pm 0.8) \cdot 10^{-3}~$\ctsper. 
A lower-background configuration, based on an exposure of 21.3~\kgy, reports a background of $(4.7 \pm 0.8) \cdot 10^{-3}~$\ctsper.

The observed lower limit on the \gess~\onbb~decay half-life is $T_{1/2}^{0\nu}>2.7 \cdot 10^{25}~$yr (90\% C.L.) with a median sensitivity for exclusion of $4.8 \cdot 10^{25}~$yr (90\% C.L.).
The half-life limit is weaker than the median sensitivity by 1$\sigma$, due to the proximity to \qbb~of an observed event at 2040~keV.

\section{The LEGEND experiment}\label{Sec:legend}

Based on the success of \gerda~and \majorana, the search for \onbb~decay in \gess~will be continued in the next years by \legend~\cite{legend} (Large Enriched Germanium Experiment for Neutrinoless \bb~Decay).
\legend~will proceed in phases towards a \onbb~decay discovery potential at a half-life beyond $10^{28}$~yr. The best technologies will be selected based
on lessons learned in \gerda~and \majorana, as well as contributions from other groups.

\begin{figure}
 \centerline{\includegraphics[width=0.56\linewidth]{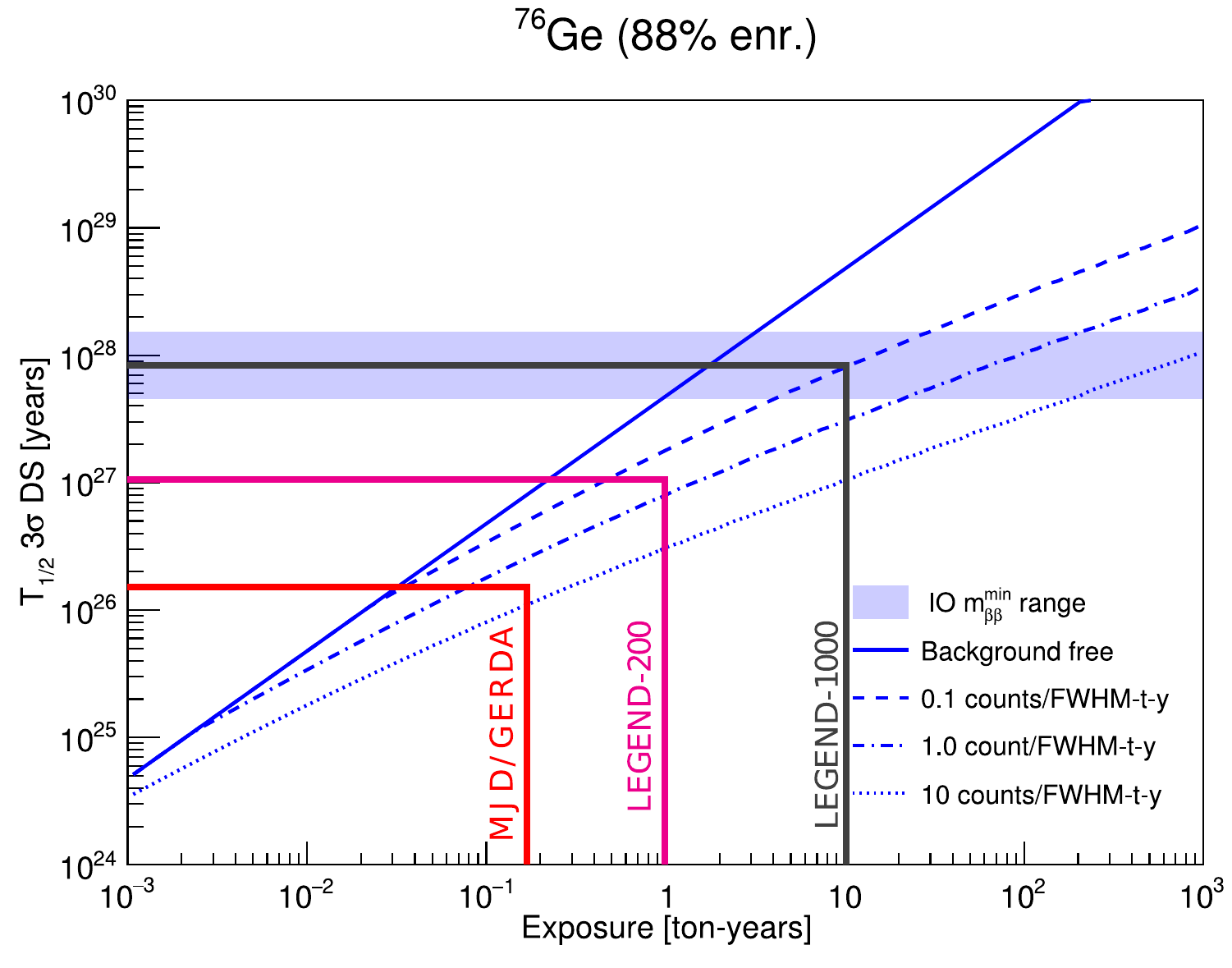}}
 \caption{\onbb~decay half-life $3\sigma$ discovery potential as function of exposure and background rate~\protect\cite{legend2018}.}
 \label{Fig:sensitivity}
\end{figure}

In its first phase, \legend-200, the existing \gerda~infrastructure at LNGS will be modified to deploy 200~kg of detectors in the cryostat. 
\legend-200 has a background goal of less than 0.6~\ctsfwhm. Achieving this background rate will allow to reach a sensitivity greater than $10^{27}$~yr with 1~\tony~of exposure. The corresponding discovery potential is shown in Fig.~\ref{Fig:sensitivity} and physics data collection is expected to begin in 2021.
Multiple techniques are already planned to achieve the background reduction required for \legend-200, such as the use of the \majorana~electroformed copper, the upgrade of the \gerda~liquid argon veto and the improvement of the front-end electronics. A crucial point is the choice of a new detector geometry, the Inverted Coaxial Point Contact (ICPC) detector~\cite{icpc}, with similar performance to the BEGe and PPC detectors and a mass as large as a coaxial detector. Five enriched ICPC detectors have been already produced and deployed in \gerda~during May 2018.

The second stage of \legend~will occur in a new infrastructure, with 1000~kg of detectors deployed. The background goal for \legend-1000 is less than 0.1~\ctsfwhm. This background reduction is necessary to achieve a \onbb~decay discovery potential at a half-life greater than $10^{28}$~yr on a reasonable timescale (see Fig.~\ref{Fig:sensitivity}). The required depth to keep cosmogenic activation backgrounds (e.g. $^{77m}$Ge) within the background budget is currently under investigation and will be a contributing factor in
the choice of site~\cite{cosmic}.

\section{Conclusions}
The latest results from \gerda~and \majorana~confirmed the high quality of the experiments and the effectiveness of background suppression techniques, consisting of selection of ultra-pure materials, detection of the LAr scintillation light and powerful pulse shape discrimination cuts.

The next generation of \onbb~decay experiments requires reduced backgrounds and additional mass. \gess~detectors have demonstrated the lowest backgrounds and best energy resolution of all competing technologies. The \legend~collaboration plans to take the best of \gerda~and \majorana~and perform additional R\&D to build a detector with \onbb~decay discovery potential at a half-life beyond $10^{28}$~yr.

\end{document}